\documentclass[aps,prd,amsfonts,nofootinbib,twocolumn]{revtex4}
   
\usepackage{graphics,epsfig}

\begin{document}

\title{Evaluating matrix elements relevant to some Lorenz violating operators}

\author{Vahagn Nazaryan} 

\email{nazaryan@jlab.org}

\affiliation{Nuclear and Particle Theory Group, Department of Physics, 
            College of William and Mary, Williamsburg, VA, 23187-8795, USA} 

\date{\today}

\begin{abstract}
Carlson, Carone and Lebed have derived the Feynman rules for a
consistent formulation of noncommutative QCD. The results they obtained 
were used to constrain the noncommutativity parameter in Lorentz 
violating noncommutative field theories. However, their constraint depended
upon an estimate of the matrix element of the quark level operator 
$\left(\not\! p-m \right)$ in a nucleon.
In this paper we calculate the matrix element of  
$\left(\not\! p-m \right)$, using a variety of confinement potential models. 
Our results are within an order of magnitude 
agreement with the estimate made by Carlson et al.
The constraints placed on the noncommutativity parameter
were very strong, and are still quite severe even if weakened by an order 
of magnitude.
\end{abstract}

\maketitle

\section{introduction}

   In the recent literature, there have been considered a number of ways to
modify the structure of space-time which can have experimental consequences.
In one of the most popular scenarios, space-time is considered to become 
noncommutative at short distance scales, with space-time 
coordinates satisfying a commutation relation of the following 
form~\cite{Hewett, Hinchliffe, Jurco, Carlson:2001sw, Godfrey, Anisimov,
Carlson:2002zb} 
\begin{equation}            
 [\hat x^{\mu},\hat x^{\nu}]=i\theta^{\mu\nu} \label{eqn_comm},
\end{equation} 
where $\hat x^{\mu}$ is a position four-vector
promoted to an operator, and $\theta^{\mu\nu}$ is a set of c-numbers 
antisymmetric in their indexes. The most striking effects of space-time non 
commutativity of the form~(\ref{eqn_comm}) are the Lorentz violating effects 
appearing in field theories, which is a consequence of $\theta^{0i}$ and 
${\varepsilon}^{ijk}{\theta}^{\imath\jmath}$ defining
preferred directions in a given Lorentz frame. 

Jur\u co, M\" oller, Schraml, Schupp and Wess~\cite{Jurco} have shown how 
to construct non-Abelian gauge theories in noncommutative spaces from a 
consistency relation. Using the same approach Carlson, Carone and 
Lebed~\cite{Carlson:2001sw}  have derived the Feynman rules for consistent 
formulation of noncommutative QCD and they have computed the most dangerous,
Lorentz-violating operator that is generated through radiative corrections.  
They have found that at the lowest order in perturbation theory, the 
formulation of noncommutative QCD that they have presented leads to 
Lorentz violating operators such as~\cite{Anisimov}
\begin{equation}
         {\theta}^{\mu\nu} \bar q{\sigma}_{\mu\nu} q,\ \ \
         \theta^{\mu\nu}\bar q\sigma_{\mu\nu} \rlap/\!D q \ \ \ {\rm and} \ \ \
              \theta^{\mu\nu}D_{\mu}\bar q\sigma_{\nu\rho}D^{\rho} q. 
          \label{eqn_LVO}
\end{equation}

In~\cite{Carlson:2001sw} the phenomenological implications of the first of 
these operators were studied in detail. Noting that contributions from the space-space
part of ${\theta}^{\mu\nu}$ make $\sigma_{\mu\nu}\theta^{\mu\nu}$ act like a 
$\vec \sigma \cdot \vec B$ interaction with $\vec B$ directly related to $\theta^{ij}$,
a limit was placed on the scale of non commutativity. One used the result of tests of 
Lorentz invariance in clock comparison experiments~\cite{Berglund}, which 
suggest that external $\vec \sigma \cdot \vec B$ like interactions are 
bounded at the $10^{-7}$Hz level or few $\times$ $10^{-31}$ GeV. 
Carlson et al.~\cite{Carlson:2001sw} concluded that 
\begin{equation} 
                       \theta{\Lambda}^2\le10^{-29}, \label{eqn_Ccon}
\end{equation}
where $\theta$ is a typical scale for elements of the matrix $\theta^{\mu\nu}$.

However, the effective Lorentz violating operator was obtained from a one loop 
correction to the quark propagator, and the operator proportional to 
$\sigma_{\mu\nu}\theta^{\mu\nu}$ also contained a factor $(\not \! p-m)$.
With $\vec B$ constant, the evaluation of $\vec \sigma \cdot \vec B$
factors out from the evaluation of $(\not \! p-m)$, and our discussion is 
focused on the later.

In~\cite{Carlson:2001sw} an ad hoc estimate was used for the matrix element of the 
operator $(\not \! p-m)$, where $m$ is the current quark mass, in getting 
limit in equation~(\ref{eqn_Ccon}). The matrix element $\langle \not\! p-m \rangle$ was 
estimated to be about ${M_N/ 3}\approx 300$ MeV, where ${M_N}$ is the nucleon 
mass. However, it has been argued that the expectation value of $(\not \! p-m)$
could be much less than this naive estimate~\cite{Pospelov}.

The aim of this paper is to calculate the matrix element of the operator 
$(\not\! p-m)$, using variety of confinement potential models, so to 
evaluate the quality of the estimate made in~\cite{Carlson:2001sw}.

The sample of potentials included four different confining potentials,
two of them purely Lorentz scalar and two of them equal mixture of scalar 
and vector. The first scalar potential is a Bag-like potential

\begin{equation}     
   V(r)=\cases {V_0,&if $r \ge R$; \cr 0,& otherwise. \cr} \phantom{nnnnnn}
    \label{eqn_3D}
\end{equation}
We also consider the one dimensional case for the nicety of the analytical 
result,
\begin{equation}
 V(z)=\cases {V_0,&if $z \le -{a\over 2} \ {\rm or} \ z \ge {a\over 2}$; \cr 0,& otherwise. \cr}
 \label{eqn_1D}
\end{equation} 
The $V_0\to \infty$ limit gives, of course, the MIT Bag model~\cite{BM:1,BM:2} 
if one does not consider the Bag energy. We will consider models of 
vector+scalar confinement next, using in one case a linear spatial potential
and on the other case a harmonic one,
\begin{equation} 
     V(r)={1\over 2}(1+\gamma^0)(V_0+\lambda r) , \label{eqn_vs_lin}
\end{equation}
or
\begin{equation}
     V(r)={1\over 2}(1+\gamma^0)Cr^2 . \label{eqn_vs_har}
\end{equation}                                

Finally we shall consider a purely scalar harmonic potential,  
\begin{equation}
                     V(r)= Cr^2 .                         \label{eqn_scl} \\
\end{equation}

In the following sections it will be assumed that the current quark mass of 
$5-10$ MeV can be neglected compared to the quark eigenenergy of several hundred MeV.
\section{Scalar Square-Well potential}

\begin{figure}
\epsfysize 4cm \epsfbox{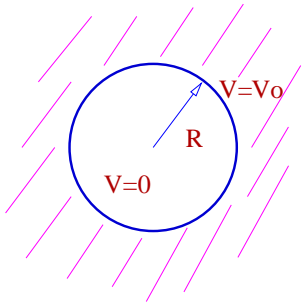}
\caption{3-D Scalar Central Confinement }  \label{fig1}
\end{figure}

For any given potential $V$, from the Dirac equation we have that
\begin{equation}
                 (\not\! p-m)\psi=V\psi  \label{eqn_Dirac},
\end{equation}
therefore
\begin{equation}
            \langle \not\! p-m \rangle = \langle V \rangle .  \label{eqn_ExpVal}
\end{equation}
In the three dimensional case, for the central potential $V(r)$ presented 
in~(\ref{eqn_3D}), the solutions of the Dirac equation for the ground state, 
with $m=0$, in two regions  $ \bf I. \ r<R, \ \text{and} \ II.\ r>R $ 
(Fig.~\ref{fig1}) have the following form
\begin{equation}
  \psi_I(r) = N_I{j_0(Er) \choose i{\bf\sigma}\cdot\hat r j_1(Er)}\chi^{(s)} ,
   \label{eqn_psi_I}
\end{equation}
\begin{equation}  
 \psi_{II}(r) = N_{II}{h_0^{(1)}(ik_0r) 
   \choose -{\bf\sigma}\cdot\hat r \sqrt{V_0-E\over V_0+E}
  h_1^{(1)}(ik_0r)}\chi^{(s)}, 
                             \label{eqn_psi_II}
\end{equation}
where $k_0=\sqrt{V_0^2-E^2}$, and $j_0, \ j_1$ are the spherical Bessel 
functions, and $h_0^{(1)}, h_1^{(1)}$ are the spherical Hankel functions
of the first kind. The ground state energy can be found from the energy eigenvalue 
equation
\begin{equation}
   j_1(ER) = j_0(ER)\left[{1+ k_0R}\over {(V_0+E)R} \right ], \label{eqn_eg}
\end{equation}  
while for $V_0 \to \infty$ the eigenvalue equation is $j_1(ER)=j_0(ER)$, as is 
familiar from the MIT Bag model~\cite{BM:1,BM:2}.  

However, we know there are long range forces between baryons. If one wants
to accommodate long range forces in this type of model, then one has to allow 
quarks to penetrate the walls of the potential well with some finite probability. 
Therefore the height of the potential, $V_0$, should be finite. A reasonable choice 
for $V_0$ and $R$ can be obtained by fitting the model parameters to obtain 
reasonable values, for example, for the mean square of charge radius of the 
nucleon $\langle r^2 \rangle$ and for the axial vector coupling constant
$g_A$. We get a good fit by choosing $R=1.12$ fm and $V_0 =3$ GeV for which we
find $\langle r^2 \rangle = 0.64\ {\rm fm^2}$ and $g_A=1.15$, as compared to experimental
values of $0.76\ {\rm fm^2}$ and 1.27 respectively~\cite{PDBook}. 
Solving~(\ref{eqn_eg}) for this choice of parameters for the ground state energy of a 
quark we find $E=348$ MeV.

Using solutions given in~(\ref{eqn_psi_I}) and~(\ref{eqn_psi_II}), we find
\begin{equation}
                \langle \not\! p-m \rangle=21\ \text{MeV}. \label{eqn_3D_expval}
\end{equation} 
Exploration of  the integrals appearing in  $\langle V(r) \rangle$, shows that  
$\langle\not\! p-m \rangle \to 0$ as $1/V_0$, when $V_0\to \infty$.

\begin{figure}
\epsfysize 4cm \epsfbox{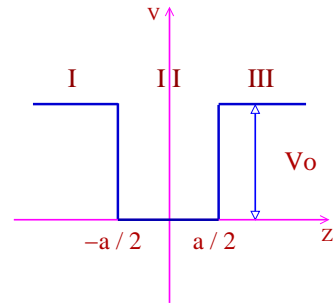}
\caption{One Dimensional Scalar Square Well Confinement}  \label{fig2}
\end{figure}

It may be of some pedagogic value to give the equivalent result for the 1D 
case~(Fig.~\ref{fig2}). The wave function for $|z| < a/2$ is just the free
solution of the Dirac equation, and the solutions for $|z| > a/2$ are
obtained from the free solution by the substitution  $E \rightarrow E-V_0$. 
We obtain 
\begin{equation}
        \langle \not\! p-m \rangle = 2V_0\int_{a\over 2}^\infty \bar\psi \ \psi\ dz 
                                       = {E \over 1+a\sqrt{ V_0^2 - E^2}} .
      \label{eqn_1D_ExpVal}
\end{equation}
One can note immediately that when the height of the potential 
$V_0\rightarrow\infty$ then $\langle \not\! p-m \rangle \rightarrow 0,$ 
unless $ a\rightarrow 0$. For the choice of parameters made above, we obtain 
\begin{equation}
                   \langle \not\! p-m \rangle= 14\ {\rm MeV}, 
      \label{eqn_1D_expval}
\end{equation}
where for the ground state energy E we have used a value of $260$ MeV, 
from the energy eigenvalue equation.  
%
\section{Scalar + Vector Linear Confinement}

Let us consider now the confinement problem of a spin 1/2 particle in a confining 
potential of the form
\begin{equation}
             V(r) = {1 \over 2}(1+\gamma^0)(V_0 + \lambda r). \label{eqn_lin_pot}
\end{equation}
This linear potential model for quark confinement was used in~\cite{Ferreira:1}
to calculate several properties of low-lying baryons. In~\cite{Ferreira:1} the 
authors assumed nonzero quark masses. 
   The straightforward modification of the wave functions for the case of 
vanishing current quark masses yields the following solution for the 
lowest energy eigenstate of the Dirac equation for the potential~(\ref{eqn_lin_pot}),
\begin{eqnarray}
  \Psi(r)&=& N{\Phi (r) \choose {{\bf\sigma}\cdot{\bf p} / E} \ \Phi (r) }\chi^{(s)}, 
   \label{eqn_Psi_lin} \\
   \Phi (r) &=& \sqrt{K\over {4\pi Ai '^2(a1)}}{1\over r} Ai(Kr+a_1)),
    \label{eqn_Phi_lin}\\ \nonumber
\end{eqnarray}
where $K=(\lambda E)^{1/3}$. The energy eigenvalue $E$ and the normalization 
constant $N$ are given in~(\ref{eqn_EN_lin})
\begin{equation}
 E = V_0 - {\lambda a_1\over K} ,\ \
 N^2 = {3E \over 4E-V_0} \label{eqn_EN_lin}.
\end{equation}
In~\cite{Ferreira:1} an analytic expression was obtained for the mean square 
charge radii of the barions and in~\cite{Ferreira:2} Ferreira obtained an analytic 
expression for the magnetic moment of the proton. We modified those expressions 
for the zero current quark mass case and used them together with the energy 
eigenvalue equation~(\ref{eqn_EN_lin}) to fit our model parameters $V_0$ 
and $\lambda$. We choose $V_0=-626$ MeV and $\lambda = 0.98$ GeV/fm
to fit $\langle r^2 \rangle$ exactly and give the closest to the data 
value of $\mu_p$, obtaining
\begin{equation}
            E = 420 {\rm MeV}, \ \ \langle r^2 \rangle = 0.76 \ {\rm fm^2} \ \
           {\rm and} \ \ \mu_p = 2.44 \ {\rm n.m.} \label{eqn_lin_rslts}
\end{equation}

For the above mentioned values of the model parameters we find that
\begin{equation}
            \langle \not\! p-m \rangle=27\ \text{MeV}. \label{eqn_lin_expval}
\end{equation}
\section{Scalar + Vector Harmonic Confinement}

Consider now potential of the form 
\begin{equation}
   V(r)={1\over 2}(1+\gamma^0)Cr^2 \label{eqn_vsh}.
\end{equation}
The solution of Dirac equation with this potential is given in~\cite{Tegen:RHOP}. 
They write the lowest energy state Dirac spinor as
\begin{equation}
   \psi(r) = {1\over \sqrt{4\pi}}{{ i f(r)/ r} 
    \choose {\bf \sigma}\cdot\hat r{ g(r)/ r}}
                    \chi^{(s)} ,  
    \label{eqn_psi_vsh}
\end{equation}
where $\chi^{(s)}$ is a Pauli spinor, with the normalization
$\int \psi^{\dagger}\psi d^3r = \int_0^\infty(f^2+g^2)dr = 1$. 
Then the upper and lower components of the solution are
\begin{eqnarray}
  f(r)&=&N\left({r\over r_0}\right)e^{-r^2/2r_0^2}, \nonumber \\
  g(r)&=&-{N\over\sqrt 3}\left({r\over r_0}\right)^2 e^{-r^2/2r_0^2},  \label{eqn_wvf_vsh} \\
  N&=&\sqrt{8/(3r_0\sqrt \pi)},\ \ r_0^2E_0^2=3,\ \ C={1\over 9}E_0^3,  \nonumber \\
  \nonumber
\end{eqnarray}
where $E_0$ is the ground state eigenenergy and $r_0$ is a state dependent 
scale parameter.

Now we can calculate the matrix element of interest,
\begin{eqnarray}
  \langle\not\! p-m \rangle &=& \int \bar{\psi}{1\over 2}(1+\gamma^0)Cr^2\psi\ d^3r 
   \nonumber \\
  &=&\int_0^\infty f(r)^2 Cr^2\ dr =\left(C\over3\right)^{1/3}={E_0\over 3}. 
    \nonumber \\ 
    \label{eqn_ExpVal_vsh}
\end{eqnarray}
So, we can see that, in case of scalar+vector confinement
of equal strengths, $\langle\not\! p-m \rangle$ is determined only 
by spin independent part of the Dirac spinor and is equal to one 
third of the ground state energy. 
In~\cite{Tegen:EPRq} it was also shown that for three massless quarks 
in their lowest 1$s$ orbit, with energy eigenvalues $E_0$ for each quark, 
the center-of-mass energy obtained with the potential~(\ref{eqn_vsh}) is just $E_0$, 
hence the nucleon mass in this model is: $M_N=2E_0$ (instead of $M_N=3E_0$ ,
as in non-relativistic and non-recoil models). Therefore, $E_0=540$ MeV and,
\begin{equation} 
             \langle \not\! p-m \rangle=180\ \rm{MeV}. \label{eqn_expval_vsh}
\end{equation}
\section{Pure Scalar Harmonic Potential}

Tegen~\cite{Tegen:RHOP} has considered scalar+vector harmonic confinement 
in calculating the weak neutron decay constant $g_A/g_V$ and found too 
small a value for $g_A/g_V$, compared to experiment. 
In~\cite{Tegen:EPRq} and ~\cite{Tegen:r2}, a pure scalar harmonic potential 
$V(r)=Cr^2$ was studied numerically, and yielded more satisfactory results for $g_A$
and for the RMS charge radius. We find that
\begin{figure}
\epsfysize 5cm \epsfbox{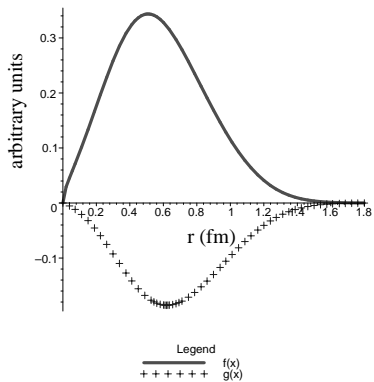}
\caption{Fit to the numerical solution of Dirac equation for 
         pure scalar harmonic confinement.
         }  \label{fig3}
\end{figure}
\begin{equation}
  \langle \not\! p-m \rangle=C\int_0^\infty r^2\left( f(r)^2 - g(r)^2\right) \ dr ,
  \label{eqn_ExpVal_sc}
\end{equation}
where $f(r)$ and $g(r)$ are defined as in eqn.~(\ref{eqn_psi_vsh}).

We have fitted the numerical solution presented graphically in ~\cite{Tegen:EPRq} with 
$C=830\ \text{MeV/fm}^2$ to calculate our integral of interest~(\ref{eqn_ExpVal_sc}). 
The fitted wave functions are presented in Fig.~\ref{fig3}, 
and as a benchmark for evaluation of the quality of the fit, we 
have calculated $\langle r^2\rangle$ and $g_A$ and obtained values $0.61$ fm and 
$1.26$ respectively, as compared to $\langle r^2\rangle=0.64$ fm and $g_A=1.26$ 
found in~\cite{Tegen:EPRq}.

Thus we obtained, without any additional tuning, the following result:
\begin{equation}
  \langle \not\! p-m \rangle = 160 {\rm MeV}. \label{eqn_expval_sc}
\end{equation}
\section{Summary}

In this paper we have calculated, for the ground state of the quark 
in a nucleon, the matrix element of the operator $(p\!\!\!\!\! \not  - m)$, 
using variety of confinement potential models, under the assumption that the 
constituent quarks obey the Dirac equation. The motivation has been to solidify 
the estimates of the non-commutativity parameter of canonical (Lorentz violating)
noncomutative QCD, where some of leading order Lorentz violating effects are 
proportional to factors of $\langle \not\! p-m \rangle$.

Interestingly, we found the following results,
\begin{equation}
   \langle \not\! p-m \rangle\! =\!  
    \cases {
            21\ \rm{MeV},&for \ 3-D scalar central pot. \cr
            27\ \rm{MeV}, &for \  scalar+vector linear pot. \cr
            180\ \ \rm{MeV}, &for \  scalar+vector harm. pot. \cr
            160\ \ \rm{MeV}, &for \  pure scalar harmonic pot.\cr
            } \label{eqn_results}
\end{equation}

We note that in the case of scalar central confinement as  considered in 
section {\bf II}, $\langle \not \!\!\!p - m  \rangle$ vanishes as $1/ V_0$ when 
$V_0 \to \infty$, but it is different from zero in general. We note also that 
the value of $\langle \not \!p - m  \rangle$ obtained for the scalar+vector linear
confinement model is close to that obtained for scalar 3D potential well.

We have also shown that  in case of scalar+vector harmonic confinement
of equal strengths, $\langle\not\!\! p-m \rangle$ is determined only 
by spin independent part of the Dirac spinor and is equal to one 
third of the ground state energy.

For pure scalar harmonic confinement of the form $V(r)=Cr^2$, 
$\langle \not\!p-m \rangle$ was obtained using a fit to the numerical solution
of the Dirac equation presented graphically in~\cite{Tegen:EPRq}, and appears to 
have a value pretty close to that obtained for the  scalar+vector harmonic 
confinement model.

Results obtained in this paper are within an order of magnitude 
agreement with the estimate made by Carlson et al.~\cite{Carlson:2001sw}.
The results obtained in ~\cite{Carlson:2001sw} were used there to constrain the
noncommutativity parameter in Lorentz violating noncommutative field theories. 
The constraints were very strong, and are still quite severe even if weakened 
by an order of magnitude.
These results may be taken as a motivation to look for space-time noncomutativity 
in Lorentz-covariant ways~\cite{Carlson:2002wj, Carlson:2001bk, Morita, Berrino}.
\begin{acknowledgments}

The author expresses his gratitude to Prof. Carl Carlson for supervision 
of the current work. The author also expresses his appreciation to Prof. 
Christopher Carone  for stimulating discussions and for careful reading of
 the manuscript.
                      
\end{acknowledgments} 
%
%
\bibliography{mat_el_bib}

\begin{thebibliography}{21}
\expandafter\ifx\csname natexlab\endcsname\relax\def\natexlab#1{#1}\fi
\expandafter\ifx\csname bibnamefont\endcsname\relax
  \def\bibnamefont#1{#1}\fi
\expandafter\ifx\csname bibfnamefont\endcsname\relax
  \def\bibfnamefont#1{#1}\fi
\expandafter\ifx\csname citenamefont\endcsname\relax
  \def\citenamefont#1{#1}\fi
\expandafter\ifx\csname url\endcsname\relax
  \def\url#1{\texttt{#1}}\fi
\expandafter\ifx\csname urlprefix\endcsname\relax\def\urlprefix{URL }\fi
\providecommand{\bibinfo}[2]{#2}
\providecommand{\eprint}[2][]{\url{#2}}

\bibitem[{\citenamefont{{J.L. Hewett, }{F.J. Petriello, and }{T.G.
  Rizzo}}(2001)}]{Hewett}
\bibinfo{author}{\bibnamefont{{J.L. Hewett, }{F.J. Petriello, and }{T.G.
  Rizzo}}}, \bibinfo{journal}{Phys. Rev. D} \textbf{\bibinfo{volume}{64}},
  \bibinfo{pages}{075012} (\bibinfo{year}{2001}).

\bibitem[{\citenamefont{{I. Hinchliffe and }{N. Kersting}}(2001)}]{Hinchliffe}
\bibinfo{author}{\bibnamefont{{I. Hinchliffe and }{N. Kersting}}},
  \bibinfo{journal}{Phys. Rev. D} \textbf{\bibinfo{volume}{64}},
  \bibinfo{pages}{116007} (\bibinfo{year}{2001}).

\bibitem[{\citenamefont{{B. Jurco, }{L. Moller, }{S. Schraml, }{P. Schupp,
  }{and J. Wess}}(2001)}]{Jurco}
\bibinfo{author}{\bibnamefont{{B. Jurco, }{L. Moller, }{S. Schraml, }{P.
  Schupp, }{and J. Wess}}}, \bibinfo{journal}{Eur. Phys J. C}
  \textbf{\bibinfo{volume}{21}}, \bibinfo{pages}{383} (\bibinfo{year}{2001}).

\bibitem[{\citenamefont{Carlson et~al.}(2001)\citenamefont{Carlson, Carone, and
  Lebed}}]{Carlson:2001sw}
\bibinfo{author}{\bibfnamefont{C.~E.} \bibnamefont{Carlson}},
  \bibinfo{author}{\bibfnamefont{C.~D.} \bibnamefont{Carone}},
  \bibnamefont{and} \bibinfo{author}{\bibfnamefont{R.~F.} \bibnamefont{Lebed}},
  \bibinfo{journal}{Phys. Lett. B} \textbf{\bibinfo{volume}{518}},
  \bibinfo{pages}{201} (\bibinfo{year}{2001}),
  \eprint[http://arXiv.org/abs]{hep-ph/0107291}.

\bibitem[{\citenamefont{{S. Godfrey and }{M.A. Doncheski}}(2002)}]{Godfrey}
\bibinfo{author}{\bibnamefont{{S. Godfrey and }{M.A. Doncheski}}},
  \bibinfo{journal}{Phys. Rev. D} \textbf{\bibinfo{volume}{65}},
  \bibinfo{pages}{015005} (\bibinfo{year}{2002}).

\bibitem[{\citenamefont{{A. Anisimov et al.}}(2002)}]{Anisimov}
\bibinfo{author}{\bibnamefont{{A. Anisimov et al.}}}, \bibinfo{journal}{Phys.
  Rev. D} \textbf{\bibinfo{volume}{65}}, \bibinfo{pages}{085032}
  (\bibinfo{year}{2002}).

\bibitem[{\citenamefont{Carlson
  et~al.}(2002{\natexlab{a}})\citenamefont{Carlson, Carone, and
  Lebed}}]{Carlson:2002zb}
\bibinfo{author}{\bibfnamefont{C.~E.} \bibnamefont{Carlson}},
  \bibinfo{author}{\bibfnamefont{C.~D.} \bibnamefont{Carone}},
  \bibnamefont{and} \bibinfo{author}{\bibfnamefont{R.~F.} \bibnamefont{Lebed}}
  (\bibinfo{year}{2002}{\natexlab{a}}),
  \eprint[http://arXiv.org/abs]{hep-ph/0209077}.

\bibitem[{\citenamefont{{C.J. Berglund et al.}}(1995)}]{Berglund}
\bibinfo{author}{\bibnamefont{{C.J. Berglund et al.}}}, \bibinfo{journal}{Phys,
  Rev. Lett.} \textbf{\bibinfo{volume}{75}}, \bibinfo{pages}{1879}
  (\bibinfo{year}{1995}).

\bibitem[{\citenamefont{{We thank M. Pospelov for discussion on this point. See
  also}{ I. Mocioiu, M. Pospelov and R. Roiban}}(2002)}]{Pospelov}
\bibinfo{author}{\bibnamefont{{We thank M. Pospelov for discussion on this
  point. See also}{ I. Mocioiu, M. Pospelov and R. Roiban}}},
  \bibinfo{journal}{Phys. Rev. D} \textbf{\bibinfo{volume}{65}},
  \bibinfo{pages}{107702} (\bibinfo{year}{2002}),
  \eprint[http://arXiv.org/abs]{hep-ph/0108136}.

\bibitem[{\citenamefont{{Chodos, A., }{R.L. Jaffe, }{K. Johnson, }{C.B Thorn,
  }{and V.F. Weisskopf}}(1974)}]{BM:1}
\bibinfo{author}{\bibnamefont{{Chodos, A., }{R.L. Jaffe, }{K. Johnson, }{C.B
  Thorn, }{and V.F. Weisskopf}}}, \bibinfo{journal}{Phys. Rev. D}
  \textbf{\bibinfo{volume}{9}}, \bibinfo{pages}{3471} (\bibinfo{year}{1974}).

\bibitem[{\citenamefont{{Chodos, A., }{R.L. Jaffe, }{K. Johnson, }{and C.B
  Thorn,}}(1974)}]{BM:2}
\bibinfo{author}{\bibnamefont{{Chodos, A., }{R.L. Jaffe, }{K. Johnson, }{and
  C.B Thorn,}}}, \bibinfo{journal}{Phys. Rev. D} \textbf{\bibinfo{volume}{10}},
  \bibinfo{pages}{2599} (\bibinfo{year}{1974}).

\bibitem[{\citenamefont{{K. Hagiwara et al.}}(2002)}]{PDBook}
\bibinfo{author}{\bibnamefont{{K. Hagiwara et al.}}}, \bibinfo{journal}{Phys.
  Rev. D} \textbf{\bibinfo{volume}{66}}, \bibinfo{pages}{010001+}
  (\bibinfo{year}{2002}), \urlprefix\url{http://pdg.lbl.gov}.

\bibitem[{\citenamefont{{P.L. Ferreira, }{J.A. Helayel, }{and N.
  Zagury}}(1980)}]{Ferreira:1}
\bibinfo{author}{\bibnamefont{{P.L. Ferreira, }{J.A. Helayel, }{and N.
  Zagury}}}, \bibinfo{journal}{IL Nuovo Cimento} \textbf{\bibinfo{volume}{55 A,
  N.2}}, \bibinfo{pages}{215} (\bibinfo{year}{1980}).

\bibitem[{\citenamefont{{P. L. Ferreira}}(1977)}]{Ferreira:2}
\bibinfo{author}{\bibnamefont{{P. L. Ferreira}}}, \bibinfo{journal}{Lett. Nuovo
  Cimento} \textbf{\bibinfo{volume}{20, N.5}}, \bibinfo{pages}{157}
  (\bibinfo{year}{1977}).

\bibitem[{\citenamefont{{R. Tegen}}(1990)}]{Tegen:RHOP}
\bibinfo{author}{\bibnamefont{{R. Tegen}}}, \bibinfo{journal}{Annals of
  Physics} \textbf{\bibinfo{volume}{197}}, \bibinfo{pages}{439}
  (\bibinfo{year}{1990}).

\bibitem[{\citenamefont{{R. Tegen, }{R. Brockmann, }{and W.
  Weise}}(1982)}]{Tegen:EPRq}
\bibinfo{author}{\bibnamefont{{R. Tegen, }{R. Brockmann, }{and W. Weise}}},
  \bibinfo{journal}{Zeitschrift fur Physik A - Atoms and Nuclei}
  \textbf{\bibinfo{volume}{307}}, \bibinfo{pages}{339} (\bibinfo{year}{1982}).

\bibitem[{\citenamefont{{R. Tegen, }{M. Schedl, }{and W.
  Weise}}(1983)}]{Tegen:r2}
\bibinfo{author}{\bibnamefont{{R. Tegen, }{M. Schedl, }{and W. Weise}}},
  \bibinfo{journal}{Phys. Lett.} \textbf{\bibinfo{volume}{125}},
  \bibinfo{pages}{9} (\bibinfo{year}{1983}).

\bibitem[{\citenamefont{Carlson
  et~al.}(2002{\natexlab{b}})\citenamefont{Carlson, Carone, and
  Zobin}}]{Carlson:2002wj}
\bibinfo{author}{\bibfnamefont{C.~E.} \bibnamefont{Carlson}},
  \bibinfo{author}{\bibfnamefont{C.~D.} \bibnamefont{Carone}},
  \bibnamefont{and} \bibinfo{author}{\bibfnamefont{N.}~\bibnamefont{Zobin}}
  (\bibinfo{year}{2002}{\natexlab{b}}),
  \eprint[http://arXiv.org/abs]{hep-th/0206035}.

\bibitem[{\citenamefont{Carlson and Carone}(2002)}]{Carlson:2001bk}
\bibinfo{author}{\bibfnamefont{C.~E.} \bibnamefont{Carlson}} \bibnamefont{and}
  \bibinfo{author}{\bibfnamefont{C.~D.} \bibnamefont{Carone}},
  \bibinfo{journal}{Phys. Rev. D} \textbf{\bibinfo{volume}{65}},
  \bibinfo{pages}{075007} (\bibinfo{year}{2002}),
  \eprint[http://arXiv.org/abs]{hep-ph/0112143}.

\bibitem[{\citenamefont{Morita}(2002)}]{Morita}
\bibinfo{author}{\bibfnamefont{K.}~\bibnamefont{Morita}}
  (\bibinfo{year}{2002}), \eprint[http://arXiv.org/abs]{hep-th/0209234 v2}.

\bibitem[{\citenamefont{Berrino et~al.}(2002)\citenamefont{Berrino, Cacciatori,
  Celi, Martucci, and Vicini}}]{Berrino}
\bibinfo{author}{\bibfnamefont{G.}~\bibnamefont{Berrino}},
  \bibinfo{author}{\bibfnamefont{S.~L.} \bibnamefont{Cacciatori}},
  \bibinfo{author}{\bibfnamefont{A.}~\bibnamefont{Celi}},
  \bibinfo{author}{\bibfnamefont{L.}~\bibnamefont{Martucci}}, \bibnamefont{and}
  \bibinfo{author}{\bibfnamefont{A.}~\bibnamefont{Vicini}}
  (\bibinfo{year}{2002}), \eprint[http://arXiv.org/abs]{hep-th/0210171}.

\end{thebibliography}
\end{document}